\documentclass[12pt,superscriptaddress,aps,prd,preprint]{revtex4}
\usepackage{amsmath}
\usepackage{amssymb}
\makeatletter
\newcommand{\bea}{\begin{eqnarray}}
\newcommand{\eea}{\end{eqnarray}}

\newcommand{\pa}{\partial}

\usepackage[dvips]{graphicx}

\begin{document}
\title{Horava-Lifshitz gravity and G\"{o}del Universe}
\author{C. Furtado, J. R. Nascimento, A. Yu. Petrov}
\affiliation{Departamento de F\'{\i}sica, Universidade Federal da Para\'{\i}ba,\\
Caixa Postal 5008, 58051-970, Jo\~ao Pessoa, Para\'{\i}ba, Brazil}
\email{furtado,jroberto,petrov@fisica.ufpb.br}
\author{A. F. Santos}
\affiliation{Instituto de F\'{\i}sica, Universidade Federal de Mato Grosso,\\
78060-900, Cuiab\'{a}, Mato Grosso, Brazil}
\email{alesandroferreira@fisica.ufmt.br}

\begin{abstract}
We prove the consistency of the G\"{o}del metric with the Horava-Lifshitz gravity whose action involves terms with $z=1$, $z=2$ and $z=3$. We show that, for different relations between the parameters of the theory, this consistency is achieved for different classes of matter, in particular, for the small cosmological constant it is achieved only for the exotic matter, that is, ghosts or phantom matter.
\end{abstract}

\maketitle

The searching for a consistent gravity theory is a key problem of the modern theoretical physics. One of the most interesting ways to solve it is based on introducing the asymmetry between time and space coordinates in such a way that a new theory would display so-called Lifshitz scale invariance $t\to l^z t$, $x^i\to l x^i$, with $z$ is an integer number called the critical exponent. As a result, a propagator of a new theory would display better ultraviolet
asymptotics which naturally improves the renormalization properties of the theory. In the case of the gravity, the corresponding approach has been developed by Horava \cite{Hor} which opened a way to formulate new gravity models, denominated as the Horava-Lifshitz (HL) gravity models and study their properties. A lot of different aspects of these theories have been recently studied (see f.e. \cite{Visser} and references therein). 

The very important issue related to the HL gravity models is a study of their classical solutions, especially, study of the problem of consistency of the known classical solutions obtained within the usual Einstein gravity with the HL gravity equations of motion which is crucial from the viewpoint of the observational verification of such gravity models. One of the principal results obtained in this way is a proof that the HL gravities admit the black hole solutions whose properties are similar to the usual Schwarzschild black holes \cite{Kon,Hat}. Another important results are the proofs that the asymptotically AdS and Friedmann-Robertson-Walker metrics \cite{HorSol} and the wormhole metrics \cite{worm} are compatible with the equations of motion of HL gravities. 

In this paper we study the compatibility of the G\"{o}del metric \cite{Godel} with the HL gravity. The key property of this metric consists in the fact that it admits the closed timelike curves (CTCs), thus, the compatibility of the G\"{o}del metric with the HL gravity implies the possibility of the CTCs within these theories. Earlier, different issues related to the G\"{o}del metric have been studied in \cite{Godelworks}.

We start with writing down the G\"{o}del metric \cite{Godel}:
\bea
ds^2=a^2\Bigl[dt^2-dx^2+\frac{1}{2}e^{2x}dy^2-dz^2+2 e^x dt\,dy\Bigl],\label{godel}
\eea
where $a$ is a positive number. Its ADM parametrization \cite{ADM} looks like:
\bea
N&=&a^2;\, N_i=(0,a^2e^x,0);\nonumber\\
g_{ij}&=&a^2\left(
\begin{array}{ccc}
-1 & 0 & 0\\
0& e^{2x}/2 & 0\\
0 & 0 & -1
\end{array}
\right).
\eea
The corresponding nontrivial Christoffel symbols are
\bea
\Gamma^{1}_{22}=\frac{1}{2}e^{2x},\, \Gamma^2_{21}=1.
\eea
The only nontrivial component of the Riemann tensor is
\bea
R_{1212}=-\frac{a^2}{2}e^{2x}.
\eea
The nontrivial components of the (spatial) Ricci tensor are
\bea
R_{11}=-1,\, R_{22}=\frac{1}{2}e^{2x}, 
\eea
and the (spatial) scalar curvature is
\bea
R=\frac{2}{a^2},
\eea
i.e. the G\"{o}del space-time, in its purely spatial part, has a constant curvature. Also, $R_{ij}R^{ij}=\frac{2}{a^4}$.
The square root of the (spatial) metric determinant is $\sqrt{g}=\frac{a^3e^x}{\sqrt{2}}$.

Also, one will need
\bea
K_{ij}=\frac{1}{2N}(\dot{g}_{ij}-\nabla_i N_j-\nabla_j N_i),
\eea
whose only nontrivial component, for the G\"{o}del metric, is
\bea
K_{12}=K_{21}=\frac{e^x}{2}.
\eea 
Here as usual in the HL case, the covariant derivatives are constructed on the base of the spatial metric only.
It is clear that $K=g^{ij}K_{ij}=0$ for this metric, and $K_{ij}K^{ij}=-\frac{1}{a^4}$. Also, we note that all covariant derivatives of the Ricci tensor vanish, $\nabla_i R_{jk}=0$, i.e. the Ricci tensor is covariantly constant. 
Therefore, the Cotton tensor defined as \cite{Kiri}
\bea
C^{ij}=\frac{\epsilon^{ikl}}{\sqrt{g}}\nabla_k(R^j_l-\frac{1}{4}R\delta^j_l).
\eea
also identically vanishes for the G\"{o}del metric.

Our aim here is to verify whether the G\"{o}del metric solves the equations of motion for the HL gravity originally derived in the paper \cite{HorSol}. So, it is time to mount these equations of motion. One must note that the nontrivial energy-momentum tensor is necessary as it is for the G\"{o}del metric in the usual case \cite{Godel}, while in \cite{HorSol} the empty space has been studied.

Here we consider the following HL gravity Lagrangian \cite{HorSol,Kiri}:
\bea
\label{lagra}
L&=&\sqrt{g}N\Big(\frac{2}{\kappa^2}(K_{ij}K^{ij}-\lambda K^2)-\frac{\kappa^2}{2w^4}C_{ij}C^{ij}+\nonumber\\&+&\frac{\kappa^2\mu}{2w^2}\frac{\epsilon^{ijk}}{\sqrt{g}}R_{il}\nabla_jR^l_k-\frac{\kappa^2\mu^2}{8}R_{ij}R^{ij}+\frac{\kappa^2\mu^2}{8(1-3\lambda)}[\frac{1-4\lambda}{4}R^2+\Lambda R-3\Lambda^2]
+{\cal L}_m\Big),
\eea
where ${\cal L}_m$ is a matter Lagrangian. The $L$ (\ref{lagra}) is a generalized Lagrangian involving terms with different $z$, i.e. $z=1$, $z=2$ and $z=3$ spatial terms. Then, we introduce the notations:
\bea
\label{def}
&&\alpha=\frac{2}{\kappa^2},\quad\,\beta=-\frac{\kappa^2}{2w^4},\quad\,\gamma=\frac{\kappa^2\mu}{2w^2},\quad\,\zeta=-\frac{\kappa^2\mu^2}{8};\nonumber\\
&&\eta=\frac{\kappa^2\mu^2(1-4\lambda)}{32(1-3\lambda)},\quad\,\xi=\frac{\kappa^2\mu^2\Lambda}{8(1-3\lambda)},\quad\,
\sigma=-\frac{3\kappa^2\mu^2\Lambda^2}{8(1-3\lambda)}.
\eea

The equations of motion look like:\\
(i) variation with respect to $N=g_{00}$:
\bea
-\alpha(K_{ij}K^{ij}-\lambda K^2)+\beta C_{ij}C^{ij}+\gamma\frac{\epsilon^{ijk}}{\sqrt{g}}R_{il}\nabla_jR^l_k+\zeta R_{ij}R^{ij}+\eta R^2+\xi R+\sigma=T^{00},
\eea
which in the case of the G\"{o}del metric reduces to
\bea
-\alpha K_{ij}K^{ij}+\zeta R_{ij}R^{ij}+\eta R^2+\xi R+\sigma=T^{00},
\eea
i.e. the corresponding component of the Einstein equation looks like
\bea 
\label{tens1}
G^{00}= \frac{\alpha}{a^4}+\frac{2\zeta}{a^4}+\frac{4\eta}{a^4}+\frac{2\xi}{a^2}+\sigma=T^{00},
\eea
which allows to determine $T^{00}$ which is thus a constant related with the parameters of the theory through the algebraic relation.

(ii) variation with respect to $N_l=g_{0l}$:
\bea
2\alpha\nabla_k(K^{kl}-\lambda Kg^{kl})=T^{0l},
\eea
which in the case of the G\"{o}del metric reduces to
$$
G^{0l}=2\alpha\nabla_kK^{kl}=T^{0l}.
$$
From here we have: $T^{01}=0$, $T^{03}=0$ and $2\alpha(\pa_1K^{12}+3K^{12})=T^{02}$, i.e. since $K^{12}=-\frac{e^{-x}}{a^4}$, we have 
\bea
\label{tens2}
G^{02}=-4\alpha\frac{e^{-x}}{a^4}=T^{02}.
\eea

(iii) the complicated system of equations corresponding to variation with respect to $g_{ij}$ whose explicit form is
\cite{Kiri}
\bea
T_{ij}&=&\frac{1}{2}\Big[(\frac{\epsilon^{mkl}}{\sqrt{g}}Q_{mi})_{;kjl}+(\frac{\epsilon^{mkl}}{\sqrt{g}}Q_m^{\phantom{k}n})_{;kin}g_{jl}-
(\frac{\epsilon^{mkl}}{\sqrt{g}}Q_{mi})_{;kn}^{;\phantom{kk}n}g_{jl}-(\frac{\epsilon^{mkl}}{\sqrt{g}}Q_{mi})_{;k}R_{jl}-\\&-&
(\frac{\epsilon^{mkl}}{\sqrt{g}}Q_{mi}R_k^n)_{;n}g_{jl}+(\frac{\epsilon^{mkl}}{\sqrt{g}}Q_{\phantom{n}m}^nR_{ki})_{;n}g_{jl}+\frac{1}{2}(\frac{\epsilon^{mkl}}{\sqrt{g}}R^n_{\phantom{n}pkl}Q^{\phantom{p}p}_m)_{;n}g_{ij}-Q_{kl}C^{kl}g_{ij}+
\nonumber\\&+&\frac{\epsilon^{mkl}}{\sqrt{g}}Q_{mi}R_{jl;k}
\Big]+\nonumber\\
&+&
\Box[N(2\eta R+\xi)]g_{ij}+N(2\eta R+\xi)R_{ij}+2N(\zeta R_{ik}R_k^j-\beta C_{ik}C^{\phantom{k}k}_{j})-[N(2\eta R+\xi)]_{;ij}+\nonumber\\&+&
\Box[N(\zeta R_{ij}+\frac{\gamma}{2}C_{ij})]-2[N(\zeta R_{ik}+\frac{\gamma}{2}C_{ik})]_{;j}^{;\phantom{j}k}+
[N(\zeta R^{kl}+\frac{\gamma}{2}C^{kl})]_{;kl}g_{ij}-\nonumber\\&-&
\frac{N}{2}[\beta C_{kl}C^{kl}+\gamma R_{kl}C^{kl}+\zeta R_{kl}R^{kl}+\eta R^2+\xi R+\sigma
]g_{ij}+\nonumber\\&+&
2\alpha N(K_{ik}K_j^k-\lambda KK_{ij})-\frac{\alpha N}{2}(K_{kl}K^{kl}-\lambda K^2)g_{ij}+\frac{\alpha}{\sqrt{g}}g_{ik}g_{jl}\frac{\partial}{\partial t}[\sqrt{g}(K^{kl}-\lambda Kg^{kl})]\nonumber\\&+&\alpha[(K_{ik}-\lambda Kg_{ik})N_j]^{;k}+\alpha[(K_{jk}-\lambda Kg_{jk})N_i]^{;k}-\alpha[(K_{ij}-\lambda Kg_{ij})N_k]^{;k}+(i\leftrightarrow j).\nonumber
\eea 
Here we note that $\frac{\epsilon^{mkl}}{\sqrt{g}}$ is a covariantly constant tensor, $(\frac{\epsilon^{mkl}}{\sqrt{g}})_{;i}=0$. Also, 
$Q_{ij}\equiv  N(\gamma R_{ij}+2\beta C_{ij})$.
In the G\"{o}del metric case, where curvature is constant, $N$ is constant, $K=0$, Cotton tensor vanishes, and none of the metric components depends on time, this equation is reduced to
\bea
T_{ij}&=&\frac{N\gamma}{2}\Big[(\frac{\epsilon^{mkl}}{\sqrt{g}}R_{mi})_{;kjl}+(\frac{\epsilon^{mkl}}{\sqrt{g}}R_m^{\phantom{k}n})_{;kin}g_{jl}-
(\frac{\epsilon^{mkl}}{\sqrt{g}}R_{mi})_{;kn}^{;\phantom{kk}n}g_{jl}-(\frac{\epsilon^{mkl}}{\sqrt{g}}R_{mi})_{;k}R_{jl}-\\&-&
(\frac{\epsilon^{mkl}}{\sqrt{g}}R_{mi}R_k^n)_{;n}g_{jl}+(\frac{\epsilon^{mkl}}{\sqrt{g}}R_{\phantom{k}m}^{n}R_{ki})_{;n}g_{jl}+\frac{1}{2}(\frac{\epsilon^{mkl}}{\sqrt{g}}R^n_{\phantom{n}pkl}R^{\phantom{p}p}_m)_{;n}g_{ij}+\frac{\epsilon^{mkl}}{\sqrt{g}}R_{mi}R_{jl;k}
\Big]+\nonumber\\
&+&
N(2\eta R+\xi)R_{ij}+2N\zeta R_{ik}R^k_j+%\nonumber\\&+&
N\zeta\Box R_{ij}-2N\zeta R_{ik;j}^{;\phantom{jjj}k}+
N\zeta R^{kl}_{\phantom{kl};kl}g_{ij}-\nonumber\\&-&
\frac{N}{2}[\zeta R_{kl}R^{kl}+\eta R^2+\xi R+\sigma
]g_{ij}+\nonumber\\&+&
2\alpha NK_{ik}K_j^k-\frac{\alpha N}{2}K_{kl}K^{kl}g_{ij}+%\nonumber\\&+&
\alpha(K_{ik}N_j)^{;k}+\alpha(K_{jk}N_i)^{;k}-\alpha(K_{ij}N_k)^{;k}+(i\leftrightarrow j).\nonumber
\eea 
This equation is still highly cumbersome. Therefore, it is convenient to elaborate it in a way similar to \cite{HorSol}:
let us denote its right-hand side (that is, the analog of the Einstein tensor) as $G_{ij}$, with we can write
\bea
\label{ein}
G_{ij}=G^{(1)}_{ij}+G^{(3)}_{ij}+G^{(4)}_{ij}+G^{(5)}_{ij}+G^{(6)}_{ij},
\eea
where (the $G^{(2)}_{ij}$, that is, the analog of $E^{(2)}_{ij}$ from \cite{HorSol}, would involve derivatives of $K=g^{ij}K_{ij}$ and hence vanishes for the G\"{o}del metric)
\bea
G^{(1)}_{ij}&=&2\alpha NK_{ik}K_j^k-\frac{\alpha N}{2}K_{kl}K^{kl}g_{ij}+%\nonumber\\&+&
\alpha(K_{ik}N_j)^{;k}+\alpha(K_{jk}N_i)^{;k}-\alpha(K_{ij}N_k)^{;k}+(i\leftrightarrow j);\nonumber\\
G^{(3)}_{ij}&=&N\xi R_{ij}-\frac{N}{2}(\xi R+\sigma)g_{ij}+(i\leftrightarrow j);\nonumber\\
G^{(4)}_{ij}&=&2N\eta RR_{ij}-\frac{N}{2}\eta R^2g_{ij}+(i\leftrightarrow j);\nonumber\\
G^{(5)}_{ij}&=&N\zeta\Box R_{ij}-2N\zeta R_{ik;j}^{;\phantom{jjj}k}+
N\zeta R^{kl}_{\phantom{kl};kl}\,g_{ij}+(i\leftrightarrow j);\nonumber\\
G^{(6)}_{ij}&=&\frac{1}{2}\frac{\epsilon^{mkl}}{\sqrt{g}}\Big[N\gamma(R_m^{\phantom{m}n})_{;kin}g_{jl}-
N\gamma(R_{mi})_{;kn}^{;\phantom{kk}n}g_{jl}-
N\gamma(R_{mi}R_k^n)_{;n}g_{jl}+N\gamma(R_{\phantom{m}m}^{n}R_{ki})_{;n}g_{jl}\Big]+\nonumber\\
&+&
2N\zeta R_{ik}R^k_j
-%\nonumber\\&-&
\frac{N}{2}\zeta R_{kl}R^{kl}g_{ij}+(i\leftrightarrow j).\nonumber
\eea
We also have took into account that the covariant derivative of $\frac{\epsilon^{mkl}}{\sqrt{g}}$ is zero due to its invariance (cf. \cite{Carroll}). 

One can see that off-diagonal terms in $G^{(1)}$, $G^{(3)}$, $G^{(4)}$ identically vanish. As for diagonal ones, we have:
\bea
&&G^{(1)}_{11}=\alpha,\quad\, G^{(1)}_{22}=-\frac{3}{2}\alpha e^{2x},\quad\, G^{(1)}_{33}=-\frac{\alpha}{2};\nonumber\\
&&G^{(3)}_{11}=a^4\sigma,\quad\, G^{(3)}_{22}=-\frac{1}{2}a^4\sigma e^{2x},\quad\, G^{(3)}_{33}=2\xi a^2+\sigma a^4;\nonumber\\
&&G^{(4)}_{11}=-4\eta,\quad\, G^{(4)}_{22}=2\eta e^{2x},\quad\, G^{(4)}_{33}=4\eta.
\eea
Then, since all first (and therefore all next) covariant derivatives of the Ricci tensor vanish for the G\"{o}del metric, we have
\bea
G^{(5)}_{ij}=0.
\eea

It remains to find $G^{(6)}$. To do it, we take into account that all covariant derivatives of the Ricci tensor vanish.
Thus, it is reduced to
\bea
G^{(6)}_{ij}&=&
4N\zeta R_{ik}R^k_j
-%\nonumber\\&-&
N\zeta R_{kl}R^{kl}g_{ij}.
\eea
Straightforward summation yields
\bea
G^{(6)}_{11}&=&-2\zeta,\nonumber\\
G^{(6)}_{22}&=&\zeta e^{2x},\nonumber\\
G^{(6)}_{33}&=&2\zeta,\nonumber\\
G^{(6)}_{12}&=&G^{(6)}_{13}=G^{(6)}_{23}=0.
\eea
As a result, we can mount the complete stress-energy tensor $G_{ij}$ (\ref{ein}). We also lower the indices of the components $G^{00}$ (\ref{tens1}) and $G^{02}$ (\ref{tens2}), obtaining thus all nontrivial components of $G_{ij}$:
\bea
\label{tens3a}
G_{00}&=&-3\Theta-\alpha-2\Sigma,\nonumber\\
G_{02}&=&-(\Theta+2\alpha+\Sigma)e^x,\nonumber\\
G_{11}&=&\alpha+\Sigma,\nonumber\\
G_{22}&=&(-\frac{3}{2}\alpha-\frac{1}{2}\Sigma)e^{2x},\nonumber\\
G_{33}&=&-\frac{\alpha}{2}+\Theta.
\eea
Here we introduced the notation $4\eta+2\zeta=4\delta$, $4\delta+2\xi a^2+\sigma a^4=\Theta$ and $\sigma a^4-4\delta=\Sigma$.
Therefore one should maintain the components of the energy-momentum tensor of matter would be equal to the components of the stress-energy tensor (\ref{tens3a}), the G\"{o}del metric will satisfy the new equations of motion. So, let us find the corresponding matter. 
 To do it, we use the expansion of the energy-momentum tensor of the matter found in \cite{Ishak}:
\bea
T_{\mu\nu}=\rho u_{\mu}u_{\nu}+p(g_{\mu\nu}-u_{\mu}u_{\nu})+u_{\mu}q_{\nu}+u_{\nu}q_{\mu}+\Pi_{\mu\nu},
\eea
where $u_{\mu}$ is the four-velocity of the fluid whose contravariant components look like $u^{\mu}=(\frac{1}{a},0,0,0)$ (and the covariant ones, thus, are $(a,0,ae^x,0)$), and the density $\rho$, pressure $p$, heat conductivity vector $q_{\mu}$ and the anisotropic stress tensor $\Pi_{\mu\nu}$ can be found, for the known $T_{\mu\nu}$, as
\bea
\rho&=&T_{\mu\nu}u^{\mu}u^{\nu};\quad\, p=\frac{1}{3}T_{\mu\nu}h^{\mu\nu},\quad\,
q^{\mu}=T_{\nu\lambda}u^{\nu}(g^{\mu\lambda}-u^{\mu}u^{\lambda});\nonumber\\ \Pi_{\mu\nu}&=&\frac{1}{2}(h_{\mu}^{\alpha}h_{\nu}^{\beta}+h_{\nu}^{\alpha}h_{\mu}^{\beta})T_{\alpha\beta}-\frac{1}{3}h_{\mu\nu}h^{\alpha\beta}T_{\alpha\beta}.
\eea
Here $h^{\mu\nu}=g^{\mu\nu}-u^{\mu}u^{\nu}$ is a projecting operator (notice that our definitions differ from those ones used in \cite{Ishak} since we use an opposite signature). Therefore, for the components of the energy-momentum tensor equal to $G_{\mu\nu}$ (\ref{tens3a}), we can find the density and the pressure:
\bea
\label{denpre}
\rho&=&-\frac{1}{a^2}(\alpha+3\Theta+2\Sigma);\nonumber\\
p&=&\frac{1}{a^2}(\Theta-\frac{7}{2}\alpha).
\eea
We note that $\alpha=\frac{2}{\kappa^2}>0$. The standard matter is characterized by positive density and pressure, i.e. the parameters of the theory satisfy the relations $\Theta-\frac{7}{2}\alpha>0$ and $\alpha+3\Theta+2\Sigma<0$. In the case $\alpha+3\Theta+2\Sigma>0$ we have ghost matter, and in the case $\Theta-\frac{7}{2}\alpha<0$ we have so-called phantom matter or X-matter. We note that all these possibilities are compatible with the definitions (\ref{def}). Actually, both ghosts \cite{Furukawa,DeFelice} and phantoms \cite{Lima} are considered now as possible candidates to solve the problem of explanation of the cosmic acceleration. 

It is interesting also to give some generic estimations: since the cosmological constant is very small, let us suggest $\Lambda\simeq 0$ in the definitions (\ref{def}). In this case we have $\Theta\simeq 4\delta$, and $\Sigma \simeq -4\delta$, while $4\delta=-\frac{\kappa^2\mu^2}{8}(\frac{1-2\lambda}{1-3\lambda})$. Therefore, we see that the signs of density and pressure given by (\ref{denpre}), which are reduced now to  
\bea
\label{restreqs}
\rho&=&-\frac{1}{a^2\kappa^2}\left[2-\frac{\kappa^4\mu^2}{8}(\frac{1-2\lambda}{1-3\lambda})\right];\nonumber\\
p&=&-\frac{1}{a^2\kappa^2}\left[7+\frac{\kappa^4\mu^2}{8}(\frac{1-2\lambda}{1-3\lambda})\right]
\eea
essentially depend on the value of $\lambda$ and on the scale of $\kappa^4\mu^2$! For example, if $\kappa^4\mu^2\ll 1$, and $\frac{1-2\lambda}{1-3\lambda}>0$, we have a fluid with negative density and pressure, that is, the ghost fluid. To classify the possible cases, we introduce the parameter $\Delta=\frac{\kappa^4\mu^2}{8}(\frac{1-2\lambda}{1-3\lambda})$. We see that $\rho>0$ at $\Delta>2$, and $p>0$ at $\Delta<-7$. Thus, we have: at $\Delta>2$ -- phantom matter or X-matter, at $-7<\Delta<2$ -- ghosts with a negative pressure, and at $\Delta<-7$ -- ghosts with a positive pressure. 

It is interesting also what interval of values of the parameters of the theory admits usual matter, that is, that one with $\rho>0$ and $p>0$. It is easy to see from (\ref{denpre}) that both these boundings are satisfied if $\Sigma<-\frac{23}{4}\alpha$, which, after taking into account the definitions (\ref{def}), yields
\bea
\frac{3(\Lambda a^2)^2+2\lambda-1}{1-3\lambda}>\frac{92}{\kappa^4\mu^2}.
\eea
Thus, for $\lambda<1/3$, the value of $\Lambda$ compatible with the usual matter is limited from below. 

To finish the discussion, it is useful to list explicit form of all components of this energy-momentum tensor, in terms of the parameters of the original action (\ref{lagra}), with $\Omega=\Lambda a^2$:
\bea
T_{00}&=&-\frac{2}{\kappa^2}+\frac{\kappa^2\mu^2}{8}(\frac{1-2\lambda}{1-3\lambda})-\frac{3\kappa^2\mu^2\Omega}{4(1-3\lambda)}+\frac{15\kappa^2\mu^2\Omega^2}{8(1-3\lambda)};\nonumber\\
T_{02}&=&-(\frac{4}{\kappa^2}+\frac{\kappa^2\mu^2\Omega}{4(1-3\lambda)}-\frac{3\kappa^2\mu^2\Omega^2}{4(1-3\lambda)})e^x;\nonumber\\
T_{11}&=&\frac{2}{\kappa^2}+\frac{\kappa^2\mu^2}{8}(\frac{1-2\lambda}{1-3\lambda})-\frac{3\kappa^2\mu^2\Omega^2}{8(1-3\lambda)};\nonumber\\
T_{22}&=&\Big[-\frac{3}{\kappa^2}-\frac{\kappa^2\mu^2}{16}(\frac{1-2\lambda}{1-3\lambda})+\frac{3\kappa^2\mu^2\Omega^2}{16(1-3\lambda)}
\Big]e^{2x};\nonumber\\
T_{33}&=&-\frac{1}{\kappa^2}-\frac{\kappa^2\mu^2}{8}(\frac{1-2\lambda}{1-3\lambda})-\frac{3\kappa^2\mu^2\Omega^2}{8(1-3\lambda)}+\frac{\kappa^2\mu^2\Omega}{4(1-3\lambda)}.
\eea

We studied the problem of the compatibility of the G\"{o}del metric with the HL gravity whose Lagrangian involves the terms corresponding to $z=1$, $z=2$ and $z=3$. Such a generic structure of the action is important since namely due to its generality (and a large enough number of free coefficients) we have the consistent system of the Einstein equations. To achieve this compatibility, we obtained, in a constructive manner, the components of the energy-momentum tensor of the matter corresponding to a G\"{o}del solution for this gravity model. We showed that, in the case of the small cosmological constant, the matter compatible with the equations of motion must be exotic. To close the paper, we note that the compatilbility of the G\"{o}del metric with the new equations of motion allows to conclude thas the CTCs are possible within the "generalized" HL gravity, at least for the matter of the form we obtained.

{\bf Acknowledgments.}
This work was partially supported by Conselho Nacional de
Desenvolvimento Cient\'\i fico e Tecnol\'ogico (CNPq) and Coordena\c c\~ao de Aperfei\c coamento de Pessoal de N\'\i vel Superior (CAPES: AUX-PE-PROCAD 579/2008). A. Yu. P. has been supported by the CNPq project No. 303461-2009/8, and A. F. S. has been supported by the CNPq project, No. 473571/2010-2.

\end{document}